\documentclass[fleqn,usenatbib]{mnras}
\usepackage{newtxtext,newtxmath}
\usepackage{float}
\usepackage{color}
\usepackage{xcolor}
\usepackage{booktabs}
\usepackage{multirow}
\usepackage{natbib}
\usepackage{comment}
\usepackage[normalem]{ulem}
\usepackage[T1]{fontenc}
\usepackage{ae,aecompl}
\usepackage{gensymb}

\usepackage{caption,subcaption}

\usepackage{graphicx}	
\usepackage{amsmath}	
\usepackage{amssymb}	
\usepackage[export]{adjustbox}[2011/08/13]

\usepackage{mathrsfs}

\newcommand{\Ms}{M$_\odot $}

\newcommand{\vir}[1]{``#1''}

\newcommand{\mob}{\texttt{MOBSE}}

\title[BHNSs in YSCs]{Dynamics of black hole $-$ neutron star binaries in young star clusters}

\author[S. Rastello et al.]{Sara Rastello$^{1,2}$, Michela Mapelli$^{1,2,3}$, Ugo N. Di Carlo$^{4,2,3}$,
 Nicola Giacobbo$^{1,2,3}$, \newauthor  Filippo Santoliquido$^{1,2}$,  Mario Spera$^{1,2,5,6}$,  Alessandro Ballone$^{1,2}$, 
 Giuliano Iorio$^{1,2}$
\\
$^{1}$Physics and Astronomy Department Galileo Galilei, University of Padova, Vicolo dell'Osservatorio 3, I-35122 Padova, Italy\\
$^{2}$INFN - Padova, Via Marzolo 8, I--35131 Padova, Italy\\
$^{3}$INAF - Osservatorio Astronomico di Padova, Vicolo dell'Osservatorio 5, I-35122 Padova, Italy\\
$^{4}$Dipartimento di Scienza e Alta Tecnologia, University of Insubria, Via Valleggio 11, I-22100 Como, Italy\\
$^{5}$Center for Interdisciplinary Exploration and Research in Astrophysics (CIERA), Evanston, IL 60208, USA
\\
$^{6}$Department of Physics \& Astronomy, Northwestern University, Evanston, IL 60208, USA
}

\date{Accepted XXX. Received YYY; in original form ZZZ}

\pubyear{2020}

\begin{document}
\label{firstpage}
\pagerange{\pageref{firstpage}--\pageref{lastpage}}
\maketitle

\begin{abstract}
  Young star clusters are likely the most common  birthplace of massive stars
across cosmic time and influence the formation of compact binaries in several
ways. Here, we simulate the formation of black hole -- neutron star binaries
(BHNSs) in young star clusters, by means of the binary population synthesis
code \texttt{MOBSE} interfaced with the $N$-body code \texttt{NBODY6++GPU}.  BHNSs
formed in young star clusters (dynamical BHNSs) are significantly more massive
than BHNSs formed from isolated binaries (isolated BHNSs):  $\sim{}40$~\% of the
dynamical BHNS mergers have total mass $>15$ M$_\odot$, while only  $\sim{}0.01$~\%  of the
isolated BHNS mergers have mass in excess of this value.  Hence, our models strongly support a dynamical formation scenario for GW190814, given its total mass $\sim{}26$ M$_\odot$, if this event is a BHNS merger.  All
our dynamical BHNSs are ejected from their parent star cluster before they
reach coalescence. Thus, a significant fraction of BHNS mergers occurring in
the field might have originated in a young star cluster.  The mass spectrum of
BHNS mergers from gravitational-wave detections will provide a clue to
differentiate between dynamical and isolated formation of BHNSs. 

\end{abstract}




\section{Introduction}
\label{intro}


The LIGO-Virgo collaboration (LVC) has detected three binary black hole (BBH)
mergers during the first observing run (O1,
\citealt{abbottO1,abbottGW150914,abbottGW151226}) and eight additional
gravitational-wave (GW) events during the second observing run (O2), seven of
them interpreted as BBHs and one associated with a binary neutron star (BNS,
\citealt{abbottGW170104,abbottGW170608,abbottGW170814,abbottO2}). The third
observing run (O3)  has just ended and already led to the publication of GW190425 (\citealt{abbottGW190425}), a compact binary coalescence with total
mass $\sim{}3.4$ M$_\odot$. This is likely the second observed BNS merger, but
has  total mass significantly larger than the known Galactic BNSs
\citep{Oezel2016}. No black hole--neutron star (BHNS) mergers were observed in
O1 and O2 \citep{abbottO2,wei2019}, and we do not know any black hole (BH) --
pulsar binary from radio observations.  While we were addressing reviewer's comments, the LVC published GW190814 \citep{abbottGW190814}, a compact binary merger with primary mass $m_1=23.2_{-1.0}^{+1.1}$ M$_\odot$ and secondary mass $m_2=2.59_{-0.09}^{+0.08}$ M$_\odot$.  The secondary object might be either the lowest mass BH or the most massive NS known to date. In the latter case, GW190814 would be the first BHNS ever observed.

BHNSs have attracted considerable interest. The observation of a tight black
hole (BH) -- pulsar binary would be a holy grail of gravity, and is one of the
main scientific goals of the Square Kilometer Array
(SKA)\footnote{https://www.skatelescope.org/}. The lack of observations of
BH--pulsar binaries with current radio facilities is not surprising:
\citet{pfahl2005} estimate  that there are  no more than one BH -- recycled
pulsar binary in the Milky Way for every $100-1000$ BNSs, of which 15 are
currently known \citep{tauris2017, 2019ApJ...876...18F}.

Similar to BNSs,  BHNS mergers might lead to the emission of short gamma-ray
bursts under some circumstances
\citep{blinnikov1984,eichler1989,paczynski1991,narayan1991, mao1994, fryer1999,
bethe1998, bethe1999, popham1999, ruffert1999,zappa2019}. The properties of
possible optical/near-infrared counterparts to BHNSs are still matter of debate
(e.g.  \citealt{fernandez2017,andreoni_2019,barbieri_2019}).

From the non-detection of BHNS mergers in  O1 and O2, the LVC inferred an upper
limit of $\sim{}610$ Gpc$^{-3}$ yr$^{-1}$ for the local merger rate density of
BHNSs \citep{abbottO2}. Theoretical predictions for the BHNS local merger rate
density $\mathcal{R}_{\rm BHNS}$ come mostly from the isolated binary scenario:
\cite{dominik2015} predict $\mathcal{R}_{\rm BHNS}\sim{}0.04-20$ Gpc$^{-3}$
yr$^{-1}$, consistent with earlier theoretical predictions
\citep{sipior2002,pfahl2005,belczynski2007,belckzynski2010,oshaughnessy2010}.
Based on the coupling between population-synthesis simulations and cosmological
simulations, \cite{mapelli2018} find  $\mathcal{R}_{\rm BHNS}\sim{}10-100$
Gpc$^{-3}$ yr$^{-1}$, consistent with \cite{artale2019} ($\mathcal{R}_{\rm
BHNS}\sim{}60$ Gpc$^{-3}$ yr$^{-1}$). Finally, recent population-synthesis
models combined with a data-driven approach yield $\mathcal{R}_{\rm
BHNS}\sim{}4-350$ Gpc$^{-3}$ yr$^{-1}$
\citep{baibhav2019,giacobbo2020,tang2020}.

Less attention has been paid to the dynamical formation of BHNSs in dense
stellar systems, such as globular clusters or young star clusters (YSCs). We
have known for a long time that dynamical exchanges are likely to
occur when the mass of the intruder is larger than the mass of one of the two
components of the binary system
\citep{hills1980,sigurdsson1993,sigurdsson1995}. Since BHs and their stellar
progenitors are among the most massive objects in a star cluster, we expect
them to be very efficient in acquiring companions through dynamical exchanges,
unless they are ejected earlier from the stellar system. Previous studies have
shown that dynamical exchanges significantly contribute to the formation of
BBHs in globular clusters (e.g. \citealt{portegieszwart2000,rodriguez2015,
rodriguez2016, rod18, rod19, fragione2018, hong2018, samsing2018, samsing2019,
choksi2019, kremer2019d,arcasedda2019}), YSCs (e.g.
\citealt{ziosi2014,mapelli2016,fujii2017,dicarlo2019b,dicarlo2019a})  and open
star clusters (e.g.
\citealt{banerjee2010,tanikawa2013,banerjee2017,banerjee2018,banerjee2019,rastello2018,kumamoto2018,kumamoto2020}).

\cite{clausen2013} studied the dynamical formation of BHNSs in globular
clusters (see also \citealt{devecchi2007,clausen2014}), finding a local merger
rate density of $\sim{}0.01-0.17$ Gpc$^{-3}$ yr$^{-1}$. While BHNSs actively
form by exchange in globular clusters, most of these systems merge in the first
$\sim{}4$ Gyr and are subsequently ejected; hence (considering that most
globular clusters formed $12$ Gyr ago) they cannot be detected by LIGO and Virgo.
Similarly, \cite{ye2020} find a local BHNS merger rate of $\mathcal {R}_{\rm
BHNS}\sim{}0.009-0.06$ Gpc$^{-3}$ yr$^{-1}$ ($\mathcal {R}_{\rm BHNS}\sim{}5.5$
Gpc$^{-3}$ yr$^{-1}$ in their extremely optimistic model) from globular
clusters.  \cite{fragione19a,fragione19b} studied BHNS mergers originated in triple systems
deriving $\mathcal {R}_{\rm BHNS}\sim{}1.9$ $10^{-4}-22$ Gpc$^{-3}$ yr$^{-1}$. 
Finally, \cite{ziosi2014} estimate an upper limit for the local
merger rate of BHNSs from YSCs $\mathcal{R}_{\rm BHNS}\le{}100$~Gpc$^{-3}$
yr$^{-1}$. The contribution of YSCs to the local merger rate of BHNSs might be
significantly higher than that of globular clusters, because the latter formed
only in the early Universe, while the former continuously form across cosmic
history. 

BHNS mergers from isolated binary evolution appear to have a relatively high
mass ratio  ($q_{\rm BHNS}=m_{\rm NS}/m_{\rm BH}\sim{}0.2-0.3$,
\citealt{giacobbo2018b,mapelli2018,mapelli2019}): the mass of the BH is
generally $m_{\rm BH}\leq{}12$ M$_\odot$, while the mass of the NS tends to be higher than that of Galactic BNSs ($m_{\rm NS}\sim{}1.5-2$
M$_\odot$). We found no estimates of the mass distribution of BHNSs from
dynamical simulations in the literature: \cite{clausen2013} just study two test
cases in which the mass of the BH is $7$ M$_\odot$ and 35 M$_\odot$,
respectively.

Here, we study the formation and evolution of BHNSs in YSCs.  YSCs are the
nursery of massive stars (which are thought to be the progenitors of BHs and
NSs). Thus, it is reasonable to expect that compact objects participate in the dynamics of
their parent YSCs, at least for few Myr. Note that the dynamical evolution of
globular clusters and YSCs are significantly different. Globular clusters have
a two-body relaxation timescale $t_{\rm rlx}$ of several hundreds Myr
\citep{spi87} and a central escape velocity of $\sim{}30$~km~s$^{-1}$,
while YSCs have $t_{\rm rlx}\sim{}10-100$~Myr and a central escape velocity of a 
few km~s$^{-1}$. Hence, while a BH can undergo a long chain of exchanges in a
globular cluster, before being ejected by dynamical or relativistic kicks,
usually the time of ejection from a YSC is much shorter ($\sim{}$ few Myr).
Moreover, the core-collapse timescale in a YSC is $\leq{}$ few Myr
\citep{mapellibressan2013}. This implies that most exchanges in globular
clusters involve BHs and neutron stars (NSs) that have already formed, while most interactions
in YSCs happen when the progenitor stars have not yet collapsed to a BH or NS,
with significant differences in the binary compact object populations
\citep{kumamoto2018,dicarlo2019a}. Furthermore, while globular clusters formed
only in the early epochs ($\sim{}8-13$ Gyr ago), YSCs represent the main
formation pathway of massive stars down to the local Universe
\citep{lada2003,portegieszwart2010}.

Simulating YSCs requires challenging $N$-body simulations, coupled with binary
population synthesis calculations. Here, we discuss a new set of 100002 direct
$N$-body simulations of YSCs, including a high binary fraction ($f_{\rm
b}=0.4$) and fractal initial conditions. We adopted the code \texttt{NBODY6++GPU}
\citep{wang2015,wang2016}, coupled with the binary population synthesis code
\texttt{MOBSE} \citep{giacobbo2018}, as described in \cite{dicarlo2019a}.

\section{Methods}

\label{met_mod}

The simulations discussed in this paper were performed with the same code as
described in \cite{dicarlo2019a}. In particular, we use the direct summation
$N$-body code \texttt{NBODY6++GPU} \citep{wang2015} coupled with the population
synthesis code \mob{} \citep{mapelli2017,giacobbo2018,giacobbo2018b,giacobbo2018c}.  \mob{}
is an upgrade of \texttt{BSE} \citep{hurley2002}, including up-to-date
prescriptions for stellar winds, for the outcome of core-collapse supernovae (SNe) and for pair instability and pulsational pair instability. Mass loss by stellar
winds  is described as $\dot{M}\propto{}Z^\beta{}$ for all massive hot stars
(O-type, B-type, Wolf-Rayet and luminous blue variable stars). The index
$\beta{}$  is defined as $\beta=0.85$ if $\Gamma_e<2/3$,
$\beta=2.45-\Gamma_e\,{}2.4$ if $2/3\leq{}\Gamma_e\leq1$, and $\beta=0.05$ if
$\Gamma_e>1$ (see \citealt{giacobbo2018} for details).

Core-collapse SNe are described as in \cite{fryer2012}. In particular,
here we adopt the rapid core-collapse SN model, which suppresses the
formation of compact objects with mass in the $2-5$ M$_\odot$ range. According
to this model, stars developing a carbon-oxygen core $m_{\rm CO}\gtrsim{}11$
M$_\odot$ collapse to a BH directly. Finally, pair instability and pulsational
pair instability are modelled as described in \cite{spera2017}  and \cite{mapelli2020}. This
implementation produces a mass gap in the BH mass spectrum between $m_{\rm
BH}\sim{}65$~M$_\odot$ and $m_{\rm BH}\sim{}120$ M$_\odot$.

We model YSCs with three different metallicities: $Z = 0.02$, $0.002$ and
$0.0002$. We ran $33334$ $N-$body simulations per each metallicity for a total of
$100002$ simulations.

YSC masses are sampled in the range $300\leq{} M_{\rm SC}/{\rm M}_\odot <1000 $ from a distribution
$dN/dM_{\rm SC}\propto M_{\rm SC}^{-2}$, reminiscent of the distribution of
YSCs in the Milky Way \citep{lada2003}.  Hence, in this work we focus on the
smallest star clusters. We will consider more massive star clusters in a
follow-up work. We choose the initial star cluster half mass radius $r_{\rm h}$
according to the Marks \& Kroupa relation \citep{marks2012}, which relates the
total mass $M_{\mathrm{SC}}$ of a SC at birth with its initial half mass radius
$r_{\rm h}$: \begin{equation} r_{\rm h}=0.10^{+0.07}_{-0.04}\,{}{\rm pc}\,{}
\left(\frac{M_{\mathrm{SC}}}{{\rm M}_{\odot}}\right)^{0.13\pm 0.04}.  \end{equation}

We generate models of star clusters that are characterized by fractal
substructures (as described in \citealt{goodwin2004}),  by using the software
\textsc{McLuster} \citep{kuepper2011}. The fractal dimension $D$ is set to be
1.6. We choose fractal initial conditions, because observations
\citep{sanchez2009,obs11,kuhn19} and hydrodynamical simulations \citep{ballone2020}
indicate that embedded star clusters have a small fractal dimension.  The YSCs are
initialised in virial equilibrium (with T/|V|=0.5).

Star masses are extracted from a Kroupa \citep{kroupa2001} initial mass
function (IMF) with $0.1 < M  < 150$ \Ms.  We did not assume any relation between star cluster mass and maximum stellar mass\footnote{\cite{weidner2006} and \cite{weidner2010} claim the existence of a relation between star cluster mass and maximum stellar mass. If we had included such relation in our initial conditions, we would have prevented the formation of the most massive stars in the smallest cluster, possibly slowing down the dissolution of these clusters.}. The orbital parameters of binary
systems are generated following
 the distributions by \cite{sana2012}. In particular, binary eccentricities $e$ are randomly drawn from a distribution $\mathcal{P}(e)\propto{}e^{-0.42}$ with $0\leq{}e<1$, while orbital periods $P$ are randomly selected from  $\mathcal{P}(\Pi)\propto{}\Pi^{-0.55}$, where $\Pi=\log_{10}(P/\mathrm{days})$ and $0.15\leq{}\Pi\leq{}6.7$. We assume an initial total binary fraction
$f_{\mathrm{bin}}=0.4$. \textsc{McLuster}  assigns the companion stars based on
mass: stars are randomly paired
by enforcing a distribution $\mathcal{P}(q)\propto{}q^{-0.1}$, where
$q=m_2/m_1$ is the ratio of the mass of the secondary to the mass of the
primary star,  
consistent with \cite{sana2012}.
All the stars more massive than 5~\Ms{} are forced to be members of binary
systems, while stars with mass $<5$~\Ms{} are randomly paired only till  we reach a
total binary fraction $f_{\mathrm{bin}}=0.4$. The result of this procedure is that the most massive stars (down to $5$ \Ms in
our case) are all members of a binary system, while the binary fraction drops
to lower values for lower star masses.  This is consistent with observational
results (e.g. \citealt{moe2017}).

The force integration in {\sc nbody6++gpu} includes a solar neighbourhood-like
static external tidal field. In particular, the potential is point-like and the
simulated star clusters are assumed to be on a circular orbit around the centre
of the Milky Way with a semi-major axis of $8\,\mathrm{kpc}$ \citep{wang2016}.
We integrate each YSC until its dissolution or for a maximum time
$t=100\,\mathrm{Myr}$.  Initially all YSCs are tidally under-filling.
As time passes, the clusters expand and the smallest ones (in the mass range $300-700$~\Ms{}) become tidally over-filling.
Only the most massive systems ($700 <M_{\rm SC}/{\rm M}_\odot{}< 1000$ ) remain
tidally under-filling for the entire simulation. Our choice of the tidal field might affect the merger rate: YSCs closer to (farther away from) the galactic centre feel stronger (weaker) tidal forces, reducing (increasing) their dissolution timescale. We will explore the effect of different tidal fields in future works.

At time $t>100$ Myr, the BHNSs that escaped {\footnote{We define escapers as those stars and binaries that
reach a distance from the centre of the YSC larger than twice the tidal radius of the cluster, as calculated by {\sc nbody6++gpu} \citep{aarsethnb7}.}} 
from the cluster evolve only due to the emission of gravitational radiation. We estimate
the coalescence timescale of these binaries with the formalism described in
\cite{peters1964}.

A summary of the initial conditions of the simulations is
reported in Table \ref{tab:table1}.

The main differences of our simulations with respect to the ones of
\cite{dicarlo2019a} are i) the mass range of star clusters (we simulate star
clusters from $300$ to $1000$ M$_\odot$, while the mass range in
\citealt{dicarlo2019a} is $1-3\times{}10^4$~M$_\odot$), ii) the metallicity
range (\citealt{dicarlo2019a} consider only $Z=0.002$, while we simulate also
$Z=0.0002$ and $0.02$), iii) the treatment of common envelope (whose parameter
$\alpha$ was set to 3 by \citealt{dicarlo2019a}, while here we adopt
$\alpha=5$, consistent with \citealt{fragos2019}), iv) the choice of the
prescription for core-collapse SNe (here we choose the rapid model by
\citealt{fryer2012}, while \citealt{dicarlo2019a} assumed the delayed model and
have BHs with mass down to $\sim{}3$ M$_\odot$), and v) the SN kick
model: in \cite{dicarlo2019a} we assumed the same model as
\cite{fryer2012}, while here we use the same prescriptions as run
CC15$\alpha{}$15 in \cite{giacobbo2018b}.  In particular, we assume that NSs receive a natal kick randomly drawn from a Maxwellian  with a one-dimensional root-mean square
$\sigma=$15~km~s$^{-1}$. BH kicks are drawn as $v_{\rm BH}=(1-f_{\rm fb})\,{}v_{\rm NS}$, where $v_{\rm NS}$ is the NS kick drawn as described above and $f_{\rm fb}$ is the fallback fraction defined in \cite{fryer12}.

In addition, we simulate a comparison sample of isolated binaries with the
stand-alone version of \mob{}.  The isolated binary sample is composed of
$3\times{}10^7$ binary systems ($10^7$ for each metallicity).  The isolated
sample is the same as run CC15$\alpha{}5$ in \cite{giacobbo2018b}.

\begin{table*}
\begin{center}
\caption{Initial conditions.}\label{tab:table1} \leavevmode
\begin{tabular}[!h]{ccccccccccc}
\hline
Set & Run number & $M_{\rm SC}$ [M$_\odot{}$] & $r_{\rm h}$ [pc] & $r_{\rm t}$ [pc] & $Z$ & $f_{\mathrm{bin}}$ & $D$ & IMF & $m_{\mathrm{min}}$ [M$_\odot{}$] & $m_{\mathrm{max}}$ [M$_\odot{}$]
\\
\hline
Z0002 & $33334$   & $3\times 10^2- 10^3$ & $0.1\times \left( M_{\mathrm{SC}}/{\rm M}_{\odot}\right)^{0.13}$ & $ 9.9 - 13.9 $ & 0.0002 & 0.4 & 1.6 & Kroupa (2001) & 0.1 & 150\\
Z002 & $33334$   & $3\times 10^2- 10^3$ & $0.1\times \left( M_{\mathrm{SC}}/{\rm M}_{\odot}\right)^{0.13}$ & $ 9.9 - 13.9 $ & 0.002 & 0.4 & 1.6 & Kroupa (2001) & 0.1 & 150\\
Z02 & $33334$  & $3\times 10^2- 10^3$ & $0.1\times \left( M_{\mathrm{SC}}/{\rm M}_{\odot}\right)^{0.13}$ & $ 9.9 - 13.9 $ & 0.02 & 0.4 & 1.6 & Kroupa (2001) & 0.1 & 150\\
\hline
\end{tabular}
\end{center}
\begin{flushleft}
\footnotesize{Column~1: Name of the simulation set.; Column~2: Number of runs;
Column~3: total mass of YSCs ($M_{\rm SC}$);
Column~4: half-mass radius ($r_{\mathrm{h}}$); 
 Column~5: tidal radius ($r_{\mathrm{t}}$) ;
Column~6: metallicity ($Z$);
Column~7: initial binary fraction ($f_{\mathrm{bin}}$);  
Column~8: fractal dimension ($D$);
Column~9: IMF;
Column~10: minimum mass of stars ($m_{\mathrm{min}}$);
Column~11: maximum mass of stars ($m_{\mathrm{max}}$).
}
\end{flushleft}
\end{table*}


\section{Results}
\label{results}

\subsection{Population of BHNSs formed in YSCs}


\begin{figure}
\centering
\includegraphics[width=0.4\textwidth]{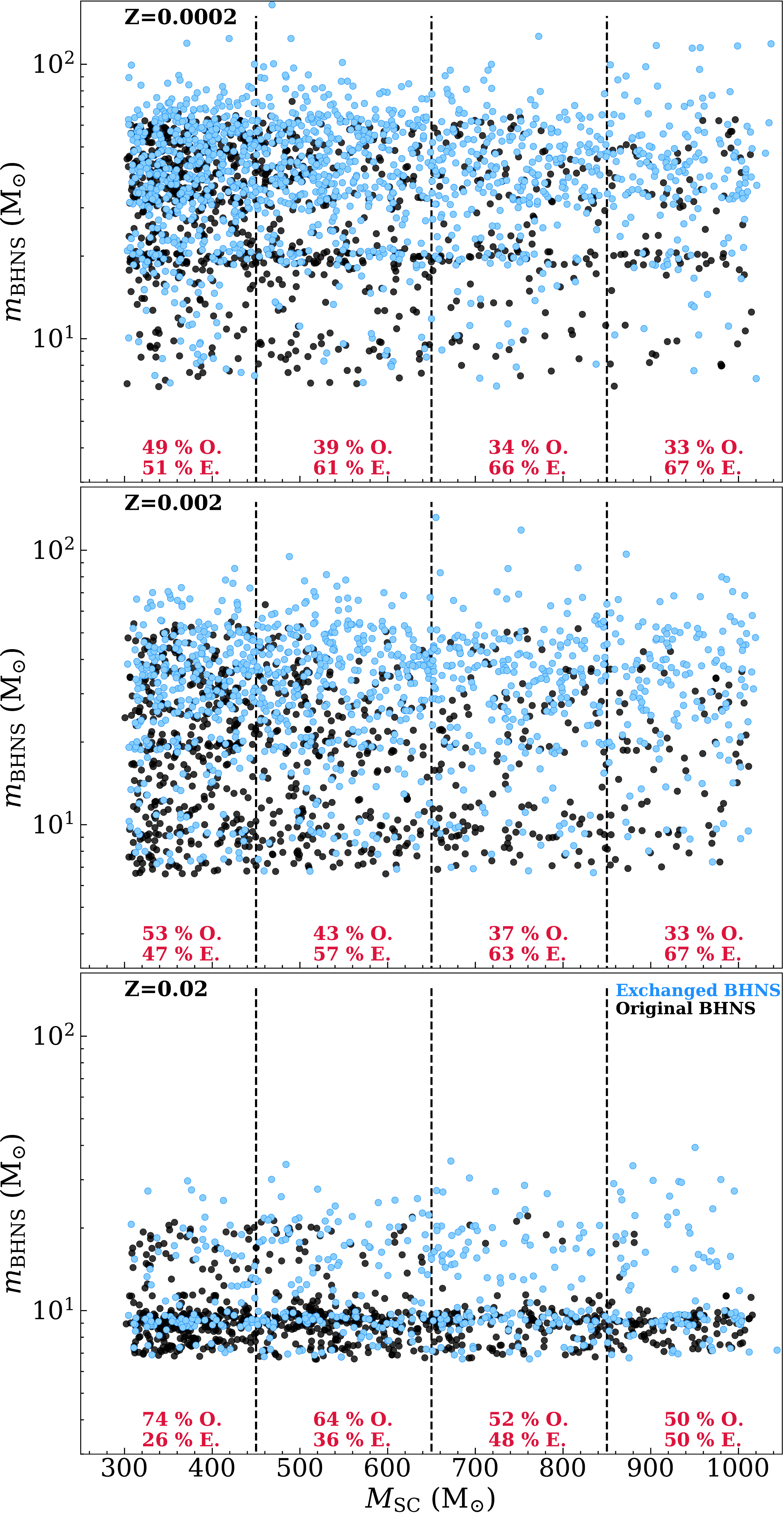}
 \caption{ Distribution of BHNS total mass
versus YSC mass for $Z=0.0002$, $0.002$ and $0.02$ (top, middle and bottom panel, respectively)
at $t=100$ Myr. Black
circles: original BHNSs; blue circles: exchanged BHNSs.  Each scatter plot is
divided into four bins of cluster mass in the range: $250<M_{\rm
SC}/$\Ms$<450$, $450<M_{\rm SC}/$\Ms$<650$, $650<M_{\rm SC}/$\Ms$<850$,
and $850<M_{\rm SC}/$\Ms$<1050$. The percentage of exchanged (E) and
original (O) BHNSs in each mass range is indicated in the bottom part of each
bin.
  }
\label{fig:bhns_formed_ysc}
\end{figure}


Figure~\ref{fig:bhns_formed_ysc} shows the population of BHNSs formed in our
$N$-body simulations at $t=100$ Myr. This sample includes both systems that
merge within a Hubble time and systems with larger orbital separation.

 The BHNSs that form from the same binary star (i.e. the stellar progenitors
of the BH and the NS were already bound in the initial conditions)
 are labelled as \vir{original BHNSs}.  The BHNSs that form through dynamical
exchanges are labelled as \vir{exchanged BHNSs}. We also consider a comparison
sample of \vir{isolated BHNSs}, which form in the field from isolated binary
evolution.  It is important to note that dynamics affects not only exchanged binaries
(which, indeed, form by dynamical encounters), but even original binaries:
close dynamical encounters shrink (or widen) the semi-major axis of a binary
star, change its orbital eccentricity and can even unbind the binary. In
particular, lighter and wider binaries (soft binaries) tend to be
widened/ionized, while massive and tight binaries (hard binaries) tend to
increase their binding energy and shrink \citep{heggie1975}.

Figure~\ref{fig:bhns_formed_ysc} shows that the percentage of exchanged
binaries increases with the total mass of the star cluster, at all considered
metallicities. The percentage of exchanged binaries is higher in metal-poor
star clusters (from $\sim{}50$~\% to $\sim{}70$~\% at $Z=0.0002-0.002$,
depending on the mass of the cluster) than in metal-rich ones (from
$\sim{}30$~\% to $\sim{}50$~\% at solar metallicity, also depending on the mass
of the cluster).

These findings can be interpreted as a result of the interplay between stellar
evolution and dynamics. According to our assumptions (in particular, to the
\citealt{marks2012} relation), our more massive star clusters are denser than
the smaller ones, hence dynamical encounters and exchanges are more common in
the former than in the latter. Moreover, BHs are generally more massive in
metal-poor clusters, hence BHs born in metal-poor star clusters are more 
efficient in acquiring companions through exchanges than BHs in metal-rich clusters.

From Fig. \ref{fig:bhns_formed_ysc} we also note that exchanged binaries are
generally more massive than original BHNSs. Dynamics leads to the formation of
more massive binaries because it allows the formation of very massive BHs
through multiple stellar collisions and dynamical exchanges allow such massive
BHs to pair with other compact objects.

Moreover, we find no evidence of correlation between the mass of the parent YSC
and the mass of the BHNSs. This is true for both original and exchanged
binaries. Each of the four bins of cluster mass shows very similar BHNS mass
distributions. The large number of BHNSs in low-mass clusters is
a direct effect of the cluster mass distribution we adopted ($dN/dM_{\rm SC}\propto{}M_{\rm SC}^{-2}$), because of which we simulated many more low-mass YSCs than high-mass YSCs. To investigate any correlation between YSC mass and number of BHNSs, we define the efficiency of BHNS formation per cluster mass as
$\eta_{f}(M_{\rm SC})=\mathcal{N}_{\rm BHNS}(M_{\rm SC})/M_{\ast}(M_{\rm SC})$,
 where $\mathcal{N}_{\rm BHNS}(M_{\rm SC})$ is the total number of BHNSs formed in YSCs with mass $M_{\rm SC}$ and $M_\ast{}(M_{\rm SC})$ is the total initial stellar mass locked in the simulated YSCs with mass $M_{\rm SC}$. Grouping our YSCs in four mass bins (the same as in Fig.~\ref{fig:bhns_formed_ysc}), we find  $\eta_{f}=7 \times 10^{-5} - 1.6\times 10^{-4}$ M$_\odot^{-1}$, varying only by a factor of two in the considered YSC mass range.

\subsection{YSCs versus isolated binaries}
\begin{figure} \centering
\includegraphics[width=0.45\textwidth]{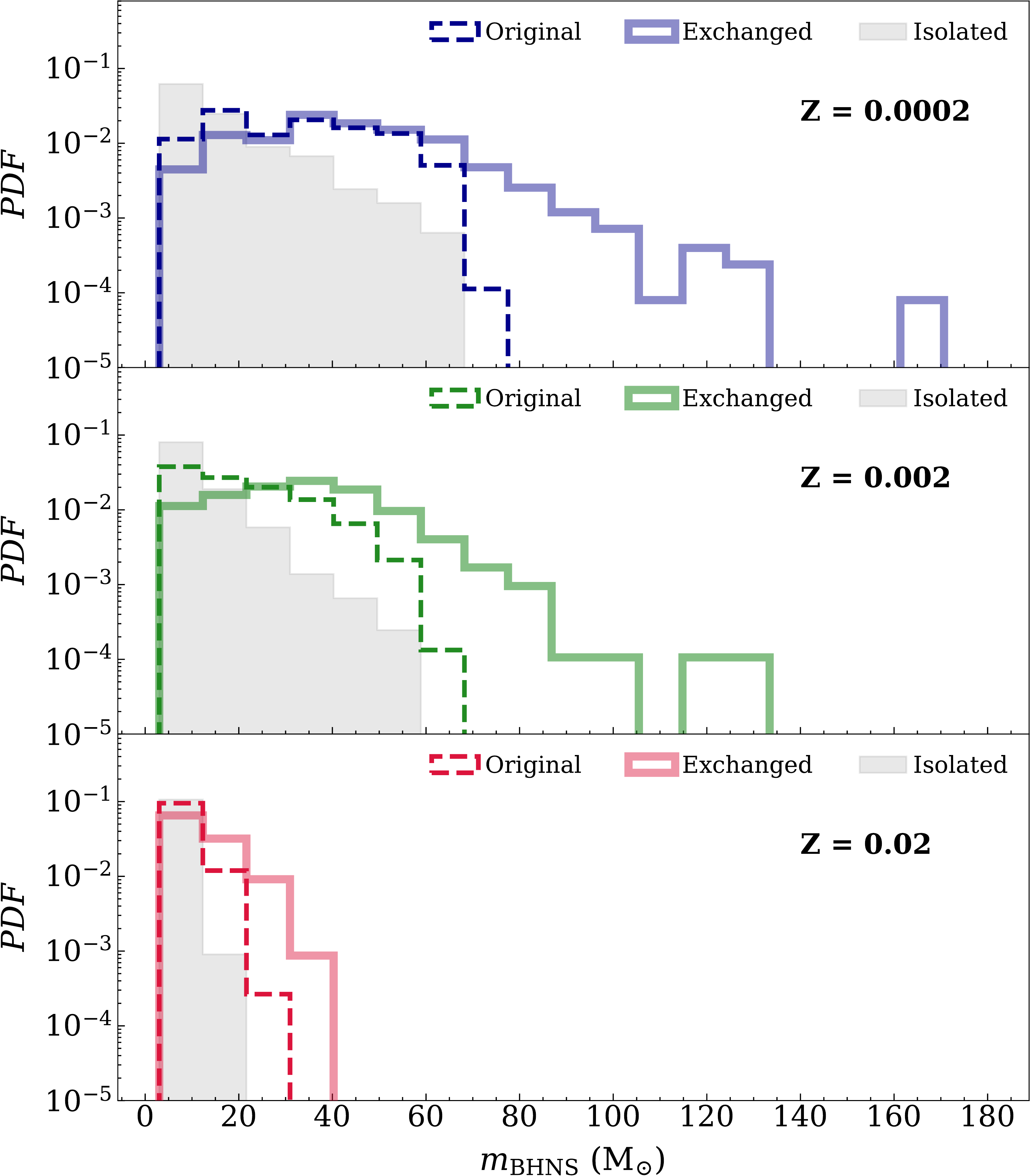}
\caption{Mass of BHNSs ($m_{\rm{BHNS}}$) formed in YSCs and in isolation.
  The panels from top to bottom refer to $Z=0.0002$ (blue), $0.002$ (green) and $0.02$ (red), respectively. The filled gray histograms refer to isolated binaries. Solid lines: exchanged BHNSs; dashed lines: original BHNSs. 
}
\label{fig:b_w_mbh_mns_ysc_field} \end{figure}

Figure~\ref{fig:b_w_mbh_mns_ysc_field} compares the mass distribution of BHNSs
formed in YSCs with that of BHNSs formed in isolation.  The
maximum mass of a BHNS, $m_{\rm BHNS,\,{}max}$, is similar in original  and
isolated BHNSs. Its value is 
 $\sim{}73$, $\sim{}63$ and $\sim{}22$ M$_\odot$  at
$Z=0.0002$ , $0.002$ and $0.02$, respectively. In contrast, the maximum BHNS
mass is significantly larger in the case of exchanged BHNSs: 
 $\sim{}164$, $\sim{}131$ and $\sim{}39$ M$_\odot$
at $Z=0.0002$, $0.002$ and $0.02$,
respectively.

 The main reason of this striking difference between exchanged binaries and the
other systems is that BHs in YSCs can form from the merger of two (or more)
stars \citep{portegieszwart2004}. In this case, the mass of the BH can be
significantly higher than the mass of a BH formed from a single star and can
even be in the pair-instability mass gap (see e.g. \citealt{dicarlo2019b} for
details).  Such massive BHs are alone at birth, but they can acquire a
companion through dynamical exchanges if they are members of a star cluster.

 Another crucial difference between isolated BHNSs and dynamical BHNSs is the
number of light systems ($m_{\rm BHNS}\lesssim{}15$~M$_\odot$) and (as a consequence) the
slope of the entire mass function. Light BHs are the most common ones in
isolated BHNSs, while their contribution is significantly smaller in dynamical
(both exchanged and original) BHNSs especially at low $Z$. This is an effect of
dynamics, because dynamical exchanges tend to suppress the lightest binaries. 
 Soft\footnote{A soft binary star is a binary star with binding energy smaller than the average kinetic energy of a star in the cluster \citep{heggie1975}.} BHNSs and their soft progenitor binary stars  tend to be disrupted by dynamical encounters during YSC evolution. Less than
5\% of such soft binaries survive till the end of the simulation. 


 About $96$\% of all the BHNSs formed in YSCs have been ejected from the
stellar system by the end of the simulations ($100$ Myr). As the metallicity of the
systems increases, the percentage of retained BHNSs decreases: $5$\% at
$Z=0.0002$, $4$\% at $Z=0.002$ and $<1$\% at $Z=0.02$. This difference is
expected, because BHNSs are generally more massive at low metallicity and thus
can be more easily retained inside the YSC.
About 60\% of the ejected BHNSs are kicked off the YSC through dynamical encounters, while the remaining 40\% have escaped because of the SN kicks. All the ejected BHNSs escape before cluster's dissolution.

\begin{figure}
\centering
\includegraphics[width=0.45\textwidth]{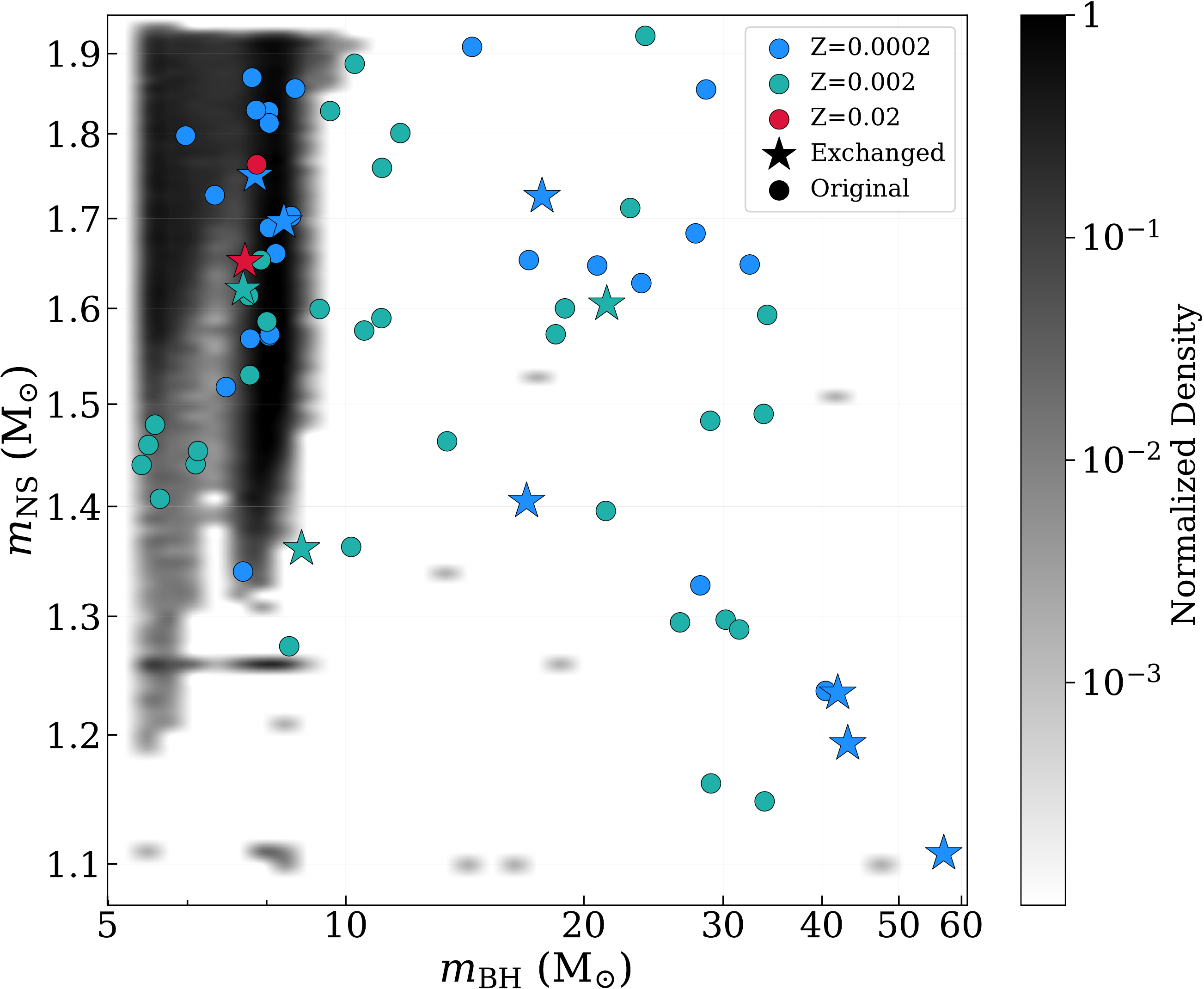}
\caption{ Mass of the BH ($m_{\rm BH}$) versus mass of the NS ($m_{\rm NS}$) of BHNS mergers.
  Circles: original BHNSs; stars: exchanged BHNSs. Blue: $Z=0.0002$; green: $Z=0.002$; red: $Z=0.02$.
Filled contours (grey colour map) indicate isolated BHNS mergers \citep{giacobbo2018} for all the three metallicities.
}
\label{fig:coal_m1m2_ysc_field}
\end{figure}

\subsection{Coalescence of BHNSs from YSCs and from isolated binaries}

In this Section, we focus on BHNSs that reach coalescence within a Hubble time
by emission of GWs.   In our dynamical simulations,  we find 69 BHNS mergers,
of which $31$, $36$ and $2$ are at metallicity
$Z=0.0002$, $0.002$ and $0.02$, respectively. Figure \ref{fig:coal_m1m2_ysc_field} shows
the mass of the NS ($m_{\rm NS}$) versus the mass of the BH ($m_{\rm BH}$) of
BHNS mergers. The majority of coalescing BHNSs from YSCs are original binaries  ($84$\%),
while the remaining ($16$\%) are exchanged binaries. The three most massive
BHs ($m_{\rm BH}>40$ M$_\odot$) in coalescing BHNSs are exchanged systems, but
even original BHNSs can host significantly massive BHs.
We estimate that $40$\% of the BHNS mergers have 
BH component with mass $m_{\rm BH}>15$ M$_\odot$.  $78$\% of such massive merging BHNS
are original, while $22$\% are exchanged binaries.
 $49$\% of the massive merging BHNSs form at $Z=0.0002$ and $51$\% at $Z=0.002$. We find no massive BHNS mergers at solar metallicity.

The mass of the NS component of BHNS mergers is always in the range
$1.1\le{}m_{\rm NS}/{\rm M}_\odot\le{}2$, but this is a consequence of the
assumed prescription for core-collapse SNe: no compact objects can form
with mass $2-5$ M$_\odot$ according to the rapid core-collapse SN model
by \cite{fryer2012}. If we had used the delayed core-collapse SN model
by the same authors, we would likely have found compact-object masses in the
$2-5$ M$_\odot$ range.


\begin{figure}
\centering
\includegraphics[width=0.45\textwidth]{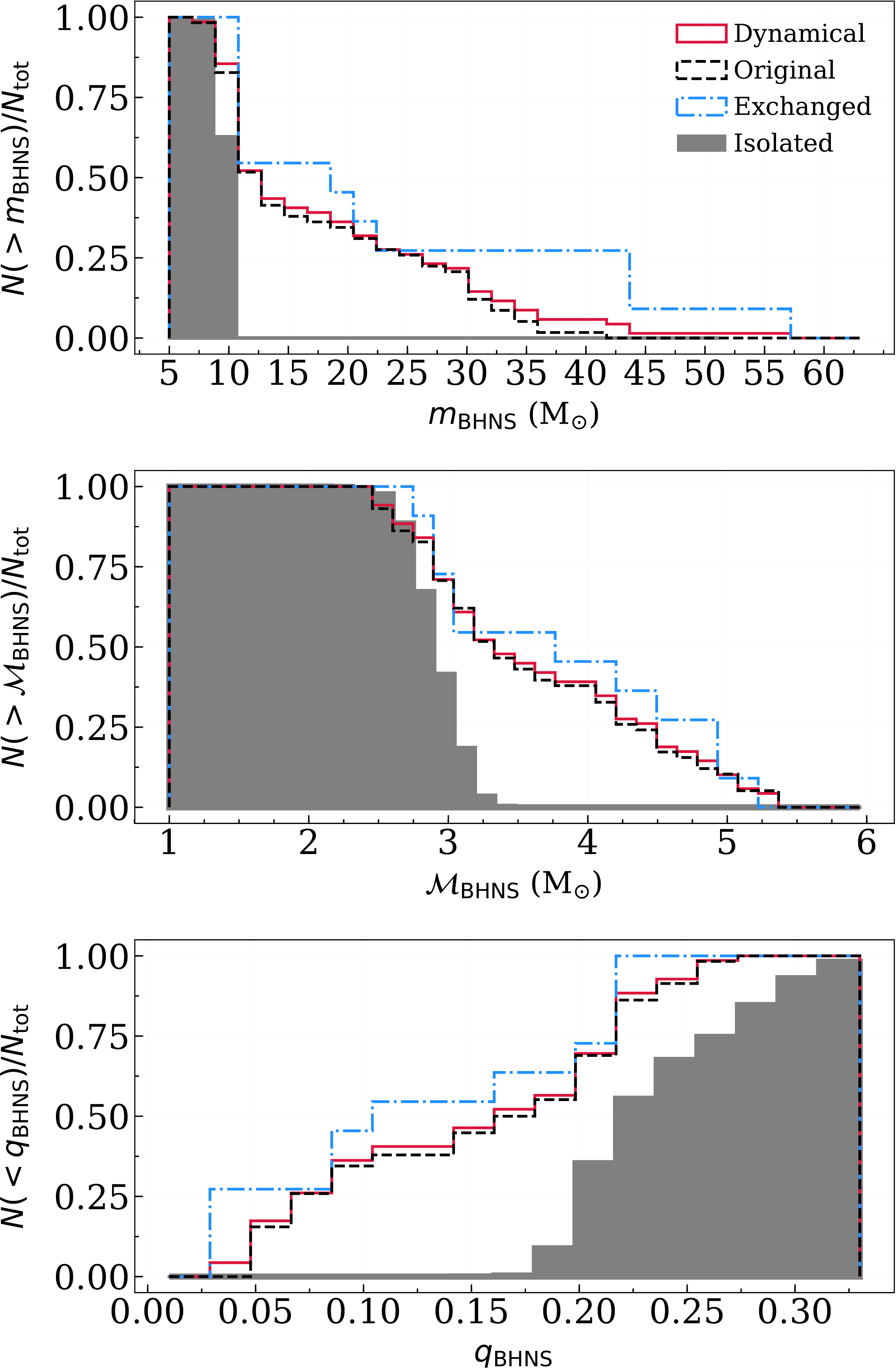}
 \caption{Cumulative distribution of total mass $m_{\rm{BHNS}}$ (top),
  chirp mass $\mathcal{M}_{\rm BHNS}$ (middle) and mass ratio $q_{\rm BHNS}$ (bottom)
 of the simulated BHNS mergers in YSCs and in isolation. Each line is normalized to the total number of mergers belonging to that specific class.
 Solid red line: all dynamical BHNS mergers (both exchanged and original BHNS mergers); dashed black line: original BHNS mergers;
 dot-dashed blue line: exchanged BHNS mergers; grey filled histograms: isolated BHNS mergers.
 The three metallicities are displayed together.} 
\label{fig:massratio}
\end{figure}

Figure~\ref{fig:massratio} shows the cumulative distribution of total mass, 
chirp mass $\mathcal{M}_{\rm BHNS}$ and mass ratio $q_{\rm BHNS}=m_{\rm NS}/m_{\rm BH}$ of BHNS
mergers.  The mass ratios are always $<0.4$, consistent with previous results
(e.g. \citealt{giacobbo2018b}). Dynamical and isolated BHNSs mergers have a
similar minimum mass ratio $q_{\rm min}\sim{}0.02$, but small mass ratios
($q<0.15$) are significantly more common in dynamical BHNSs than in isolated
BHNSs.

 The chirp masses of dynamical BHNS mergers have a much broader range of values
than the isolated systems: the former extend from
$\mathcal{M}_{\rm BHNS}\sim2.7$~M$_\odot$ to $\mathcal{M}_{\rm BHNS}\sim{}5.4$~M$_\odot$ (with
negligible differences between original and exchanged BHNSs), while the latter 
are more concentrated in the $2\leq{}\mathcal{M}_{\rm BHNS}/{\rm M}_\odot\leq{}4$ range
with a tail at higher chirp mass.

In summary, massive BH components ($m_{\rm BH}>15$~M$_\odot$) are significantly
more common in dynamical BHNS mergers than in isolated BHNS mergers. Massive
BHs are very common not only in exchanged BHNSs, but even in original BHNSs.
This result is  easy to understand in the case of exchanged BHNSs (these can
host more massive BHs born from single star evolution or from previous stellar
mergers), but is trickier to grasp for original BHNSs. The higher fraction of
massive BHs in original BHNSs with respect to isolated BHNSs comes from an
interplay between binary evolution and dynamics. As we have already discussed
in \cite{giacobbo2018b},
the most massive BHs in our models come from metal-poor stars with mass
$\sim{}60-80$ M$_\odot$. These stars develop very large radii (hundreds to
thousands of solar radii, \citealt{spera2018}) during their giant phase. If the
initial orbital separation of the binary star was smaller than these large
radii, the binary star merges before giving birth to a BHNS. For larger
orbital separation, the binary star undergoes Roche lobe overflow, which tends
to equalize the final mass of the two compact objects: the final BHNS might
merge by GW emission, but the mass of the BH is significantly smaller than
expected from single star evolution because of mass transfer and envelope
removal. Finally, if the binary is too large to undergo Roche lobe overflow
(orbital separation $a\gtrsim{}10^3$ R$_\odot$, \citealt{spera2018}), the mass
of the BH in the final BHNS is the same as expected from single star evolution
(i.e. $50-65$ \Ms for a metal-poor progenitor with zero-age main-sequence mass
$\sim{}60-80$ \Ms), but, if the binary is isolated, the final orbital
separation is too large to lead to coalescence by GW emission. In a dynamical
environment such as an YSC,  such original BHNSs with a massive BH component
can shrink by dynamical encounters and might be able to merge by GW emission.

Finally,  all the dynamical BHNS mergers happen after the binary was ejected
from the YSC.  About $70\%$ of them are ejected via dynamical encounters, while the remaining $\sim{}30$\%  is kicked off by SN kicks.

This is a crucial result because it means that the vast majority
of BHNSs born in YSCs are field binaries by the time of their merger. The
population of BHNS mergers in the field is then the result of a mixture between
genuine isolated binaries and dynamical systems previously ejected from their
parent star cluster.

\subsection{Merger efficiency and rate} \label{sec:rate}

We estimate the merger efficiency,
$\eta{}(Z)$, defined as the number of mergers $\mathcal{N}_{{\rm TOT}}(Z)$ within a Hubble
time, divided by the total initial stellar mass of the YSCs at a given
metallicity: $\eta{}(Z) = \mathcal{N}_{{\rm TOT}}(Z)/M_{\ast}(Z)$, where 
$M_\ast(Z)=\sum{}_i\,{}M_{\rm SC,\,{}i}(Z)$.  The merger efficiency is a good
proxy for the merger rate, because it does not depend on assumptions on star 
formation rate, metallicity evolution and delay time (apart from
its integrated value). Table~\ref{tab:merg_eff} shows the merger efficiency
for BHNSs from YSCs ($\eta_{\rm YSC}$) and from isolated binaries ($\eta_{\rm
IB}$) at different metallicities. In metal-poor systems ($Z=0.0002$, $0.002$),
the merger efficiency of BHNSs from small YSCs is about a factor of $10$ lower
than the BHNS merger efficiency from isolated binaries. In contrast, at solar
metallicity ($Z=0.02$) the BHNS merger efficiency associated with YSCs is about
a factor of 40 higher than the BHNS merger efficiency from isolated binaries.

This result can be interpreted as follows. In metal-poor environments, where
very massive BHs can form ($m_{\rm BH}\ge{}30$ M$_\odot$), exchanges favour the
formation of BBHs and suppress the formation of BHNSs, because NSs are much
lighter than BHs. In metal-rich environments, where BHs are rather light,
dynamics enhances the merger rate of BHNSs.

\begin{table}
\centering
\begin{tabular}{@{}ccc@{}}
\toprule
$Z$    &   $\eta_{\rm YSC}$   & $\eta_{\rm IB}$   \\
       &   $[{\rm M}_\odot^{-1}]$   & $[{\rm M}_\odot^{-1}]$    \\ \midrule
0.0002 &   $1.8\times{}10^{-6}$   & $1.7\times{}10^{-5}$      \\
0.002  &   $2.1\times{}10^{-6}$   & $2.4\times{}10^{-5}$      \\
0.02   &   $1.1\times{}10^{-7}$   & $2.6\times{}10^{-9}$       \\ \bottomrule
\end{tabular}
 \caption{Merger efficiency of BHNSs from YSCs and from isolated binaries. Column 1: metallicity $Z$; column 2: BHNS merger efficiency for YSCs $\eta_{\rm YSC}$;
 column 3: BHNS merger efficiency for isolated binaries $\eta_{\rm IB}$, from Giacobbo \&{} Mapelli (2018).}
\label{tab:merg_eff}
\end{table}

 From the merger efficiency, we can estimate the merger rate density in the
local Universe as  described in \cite{santoliquido2020}:

\begin{eqnarray}
\label{eq:rate}
   \mathcal{R}_{\text{BHNS}} = \frac{1}{t_{\rm lb}(z_{\text{loc}})}\int_{z_{\rm max}}^{z_{\text{loc}}}\psi(z')\,{}\frac{{\rm d}t_{\rm lb}}{{\rm d}z'}\,{}{\rm d}z' \times{}\nonumber{}
   \\
   \int_{Z_{\rm min}(z')}^{Z_{\rm max}(z')}\eta{}(Z)\,{}\mathcal{F}(z',z_{\text{loc}}, Z)\,{}{\rm d}Z,
\end{eqnarray}
where $t_{\rm lb}(z_{\text{loc}})$ is the look-back time evaluated in the local universe ($z_{\text{loc}}\leq 0.1$), $\psi(z')$ is the cosmic SFR density at redshift $z'$ (from \citealt{madau2017}), $Z_{\rm min}(z')$ and $Z_{\rm max}(z')$ are the minimum and maximum metallicity of stars formed at redshift $z'$  
and $\mathcal{F}(z',z_{\text{loc}}, Z)$ is the fraction of BHNSs that form at redshift $z'$ from stars with metallicity $Z$ and merge at redshift $z_{\text{loc}}$ normalized to all BHNSs that form from stars with metallicity $Z$. To calculate the lookback time $t_{\rm lb}$ we take the cosmological parameters ($H_{0}$, $\Omega_{\rm M}$ and $\Omega_{\Lambda}$)  from \cite{ade2016}. We integrate equation~\ref{eq:rate} up to redshift $z_{\rm max}=15$, which we assume to be the epoch of formation of the first stars.

From equation~\ref{eq:rate}, we obtain a local merger rate density 
$\mathcal{R}_{\rm BHNS}\sim{}28$ Gpc$^{-3}$~yr$^{-1}$, by assuming that all the
cosmic star formation rate occurs in YSCs like the ones we
simulated in this paper. For the isolated binaries, we find $\mathcal{R}_{\rm BHNS}\sim{}49$ Gpc$^{-3}$~yr$^{-1}$ \citep{santoliquido2020}.

 The models presented in this work assume low natal kicks  for NSs,  which  are  in  tension  with  the proper motions of some Galactic young pulsars \citep{giacobbo2018b}.  We recently proposed a new model for natal  kicks \citep{giacobbo2020} that  can  reproduce  the  proper  motions  of Galactic  pulsars  and  gives  a  value  for  the  merger rate  close to the one presented in this study for isolated BHNSs.   As  a  result,  we  do  not  expect  significant  differences  in  the  merger-rate density  between  the  model  adopted  in this work and the one proposed by \cite{giacobbo2020}.

\subsection{GW190814}
 While we were addressing the reviewer's comments, the LVC published the discovery of GW190814 \citep{abbottGW190814}, a binary compact object merger with a total mass of 25.8$_{-0.9}^{+1.0}$ M$_\odot$ and mass ratio $q=0.112_{-0.009}^{+0.008}$. GW190814 might be either a BBH or a BHNS, depending on the nature of the secondary component, which has mass $m_2=2.59_{-0.09}^{+0.08}$ M$_\odot$. In our models, dynamical BHNSs with total mass $\sim{}26$ M$_\odot$ are rather common, while isolated BHNSs with such high total mass and low mass ratio are extremely rare. 

In our simulations, we do not have any NS with mass $>2$ M$_\odot$, but this is an effect of our assumptions for the core-collapse SN model: we use the rapid core-collapse SN model by \cite{fryer2012}, which was designed to enforce a mass gap between 2 and 5 M$_\odot$. If we had run our simulations with, e.g., the delayed core-collapse SN model by the same authors, the gap between 2 and 5 M$_\odot$ would have disappeared.

The local merger rate density of GW190814-like systems inferred from the LVC is $\sim{}7_{-6}^{+16}$ Gpc$^{-3}$ yr$^{-1}$ \citep{abbottGW190814}. With the method described in Section~\ref{sec:rate},  we estimate a local merger rate density  $\sim{}0$ and $\sim{} 8_{-4}^{+4}$  Gpc$^{-3}$ yr$^{-1}$ for isolated  and dynamical GW190814-like systems. Here, we define GW190814-like systems as all simulated BHNS mergers with total mass $20-30$ M$_\odot$.

If we interpret GW190814 as a BBH, rather than a BHNS, we still expect that such extreme mass ratio is rather prohibitive for isolated binary evolution, while it is common in dynamical BBHs (see e.g. the companion papers by \citealt{dicarlo2019a,dicarlo2020}). Hence, our models strongly support a dynamical formation for GW190814 in a YSC.

\section{Conclusions} 
We have studied the formation of BHNSs in 100002 low
mass ($300-1000$ \Ms) young star clusters (YSCs) by means of direct $N$-body
simulations coupled with binary population synthesis. We have used a version of
\texttt{NBODY6++GPU} \citep{wang2015} interfaced with our population-synthesis
code \texttt{MOBSE} \citep{giacobbo2018}, as described in \cite{dicarlo2019a}.
Very few studies address the dynamics of BHNSs
\citep{devecchi2007,clausen2013,fragione19a,fragione19b,ye2020} and none of them focus on YSCs. 
YSCs are generally less massive than globular clusters and
short-lived, but they form all the time across cosmic history: YSCs are the
main nursery of stars in the local Universe. Moreover, none of the previous
works investigate the impact of star cluster dynamics on the mass of BHNSs.

We find that BHNSs formed in YSCs are significantly more massive than BHNSs
formed from isolated binary evolution.  At low metallicity, the mass of the BH
component in a BHNS can  reach $\sim{}160$ M$_\odot$ in YSCs and $\sim{}65$
M$_\odot$ in isolated binaries, respectively. If we focus on dynamical BHNSs
that merge within a Hubble time by GW emission, the vast majority of BHNSs in
isolated binaries ($>99$~\%) have mass $m_{\rm BHNS}\leq{}15$ M$_\odot$, while
$\sim{}40$~\% of BHNSs in YSCs have mass $m_{\rm BHNS}>15$ M$_\odot$.
 The mass range of dynamical BHNSs  in our models strongly supports a dynamical formation for GW190814 \citep{abbottGW190814}.
Interestingly, not only the exchanged BHNSs (i.e. BHNS systems formed by
dynamical exchanges) but also original BHNSs in YSCs (i.e. BHNS systems that
form in a YSC from the evolution of a primordial binary star) are significantly
more massive than BHNSs formed in isolation. This indicates that dynamical
hardening is important for BHNSs in YSCs.

Our simulations do not include compact object spins, because of the large
theoretical uncertainties about their magnitude. On the other hand, we expect
that dynamical encounters completely randomize the direction of the spins, at
least in the case of exchanged binaries \citep{bouffanais2019}. This implies
that our dynamical BHNSs have non-zero components of the spin in the orbital
plane, showing precession. Binaries with non-aligned spins and small mass ratio
$q_{\rm BHNS}=m_{\rm NS}/m_{\rm BH}$ are  not expected to be accompanied by bright
electromagnetic counterparts (e.g. \citealt{zappa2019}).

 All the BHNSs formed in YSCs merge after they were
ejected from their parent star cluster. This implies that a large fraction of
BHNS mergers in the field might have formed in YSCs.

In metal-poor YSCs ($Z=0.0002$, $0.002$), the BHNS merger efficiency
of YSCs is a factor of $10$ lower than that of isolated binaries. In contrast, at
solar metallicity ($Z=0.02$) the BHNS merger efficiency of YSCs is a factor of
40 higher than the BHNS merger efficiency of isolated binaries: dynamics
triggers a significant number of BHNS mergers at solar metallicity and reduces
the differences between metal-poor and metal-rich environments.

 Finally, we estimate a local merger rate density $\mathcal{R}_{\rm
BHNS}\sim{}28$~Gpc$^{-3}$~yr$^{-1}$, similar to recent estimates from
isolated binary evolution
\citep{mapelli2018,artale2019,baibhav2019,giacobbo2020,tang2020,santoliquido2020} and  below the
upper limit inferred from the first and the second observing runs of LIGO and
Virgo \citep{abbottO2}. Hence, a large fraction of BHNS mergers occurring in
the field might have originated in a YSC. We expect that the mass spectrum of
BHNS mergers from GW  detections will provide a clue to differentiate between
dynamical and isolated formation of BHNSs.

\section*{Acknowledgements} 
We thank the referee, Dr Mirek Giersz, for his useful comments, which helped us improving this work. MM, AB, GI, NG and SR acknowledge financial support by the European Research Council for the ERC Consolidator grant DEMOBLACK, under contract no. 770017. MS acknowledges funding from the European Union's Horizon 2020 research and innovation programme under the Marie-Sklodowska-Curie grant agreement No. 794393. UNDC acknowledges financial support from Universit\`a degli Studi dell'Insubria through a Cycle 33rd PhD grant. We thank G. Costa and M. Pasquato for interesting and stimulating discussion. We are grateful to L. Wang for his helpful support on {\sc nbody6++gpu}.
All the $N$-body simulations discussed in this paper were performed with the supercomputer DEMOBLACK at the Physics and Astronomy department \vir{G. Galilei} of the University of Padova, equipped with 192 cores and 8 NVIDIA Tesla V100 GPUs.

\bibliographystyle{mnras}
\bibliography{rasetal}

\end{document}